\documentclass[aps,prl,groupedaddress,amsmath,amssymb,twocolumn]{revtex4}

\usepackage{amssymb}
\usepackage{amsmath}
\usepackage{graphicx}
\usepackage{textcomp}
\usepackage{color}
\usepackage{amsfonts}
\usepackage{epsfig}
\usepackage{hyperref}

\usepackage{pdfpages}

\newcommand{\nabisb}{Na$_3$Bi$_{1-x}$Sb$_{x}$}

\begin{document}
\title{Topological tuning in three-dimensional Dirac semimetals}

\author{Awadhesh Narayan}
\email{narayaa@tcd.ie}
\affiliation{School of Physics and CRANN, Trinity College, Dublin 2, Ireland}

\author{Domenico Di Sante}
\email{domenico.disante@aquila.infn.it}
\affiliation{Consiglio Nazionale delle Ricerche (CNR-SPIN), Via Vetoio, L'Aquila, Italy} 
\affiliation{Department of Physical and Chemical Sciences, University of L'Aquila, Via Vetoio 10, I-67010 L'Aquila, Italy}

\author{Silvia Picozzi} 
\affiliation{Consiglio Nazionale delle Ricerche (CNR-SPIN), Via Vetoio, L'Aquila, Italy}

\author{Stefano Sanvito}
\affiliation{School of Physics, CRANN and AMBER, Trinity College, Dublin 2, Ireland}
\date{\today}

\begin{abstract}
We study with first-principles methods the interplay between bulk and surface
Dirac fermions in three dimensional Dirac semimetals. By combining density
functional theory with the coherent potential approximation, we reveal a
topological phase transition in \nabisb and Cd$_3$[As$_{1-x}$P$_x$]$_2$ alloys,
where the material goes from a Dirac semimetal to a trivial insulator upon
changing Sb or P concentrations. Tuning the composition allows us to engineer the
position of the bulk Dirac points in reciprocal space. Interestingly, the phase
transition coincides with the reversal of the band ordering between the
conduction and valence bands.
\end{abstract}

\maketitle

\textit{Introduction}. In recent years an ever growing attention has been devoted to Dirac fermions, both in two as well 
as in three dimensions. The fabrication of graphene and topological insulators has motivated a surge of investigations 
in this field~\cite{review-geim,review-kane,review-zhang}. Acting as a possible bridge between the two, recently Dirac 
semimetals in three dimensions were theoretically proposed~\cite{kane-semimetal}. Using first-principles calculations, 
Wang and co-workers predicted sodium bismuthate (Na$_3$Bi) and cadmium arsenide (Cd$_3$As$_2$) to be three 
dimensional Dirac semimetals~\cite{fang-na3bi,fang-cd3as2}. Their experimental realization has not been far behind 
and the prediction verified by means of angle resolved photoemission measurements in a remarkably rapid flurry of 
activity by a number of groups~\cite{chen-na3bi,hasan-na3bi,chen-cd3as2,hasan-cd3as2,cava-cd3as2}. Interestingly, 
a Dirac semimetal state was also found in zinc blende compounds~\cite{potemski-znb}. Apart from hosting a bulk Dirac 
cone, both Na$_3$Bi and Cd$_3$As$_2$ also show a band inversion at the center of the Brillouin zone (BZ). This means 
that they exhibit a surface Dirac spectrum when confined to a slab geometry, analogously to conventional topological
insulators~\cite{zhang-bi2se3}. Given their unique electronic structure, these class of compounds opens up an exciting 
platform to study topological phase transitions, interweaving two and three dimensional Dirac states.

In this Letter we study the interplay of surface and bulk Dirac states by
using first-principles density functional theory (DFT) and
\textit{ab-initio}-derived tight-binding models. Based on our first-principles
calculations, we predict that the bulk Dirac cone of Na$_3$Bi is formed only for
films with a thickness larger than 90~nm, while the surface Dirac state,
originating from a bulk band inversion, becomes gapless for films with a
thickness as small as 4.5~nm, up to an energy resolution of $\approx$ 3 meV (a
resolution accessible by the most recent state-of-the-art
spectrometers~\cite{damascelli}). Furthermore, by employing the coherent
potential approximation joint with DFT, we uncover a topological phase
transition in the \nabisb~and Cd$_3$[As$_{1-x}$P$_x$]$_2$ alloys. We propose a means to engineer the $k$-space
position of the bulk Dirac point by changing the Sb or P concentrations. At the
critical Sb (P) concentration of $\approx$ 50\% ($\approx$ 10\%), this crossing reaches the BZ
center, meeting its time-reversed partner, whereupon they annihilate and render
the bulk gapped. This topological phase transition is accompanied by a
simultaneous loss of the inverted band character. Beyond the critical Sb (P)
concentration, the alloy is adiabatically connected to the topologically trivial
Na$_3$Sb (Cd$_3$P$_2$).
\begin{figure}[b]
\begin{center}
  \includegraphics[scale=0.30]{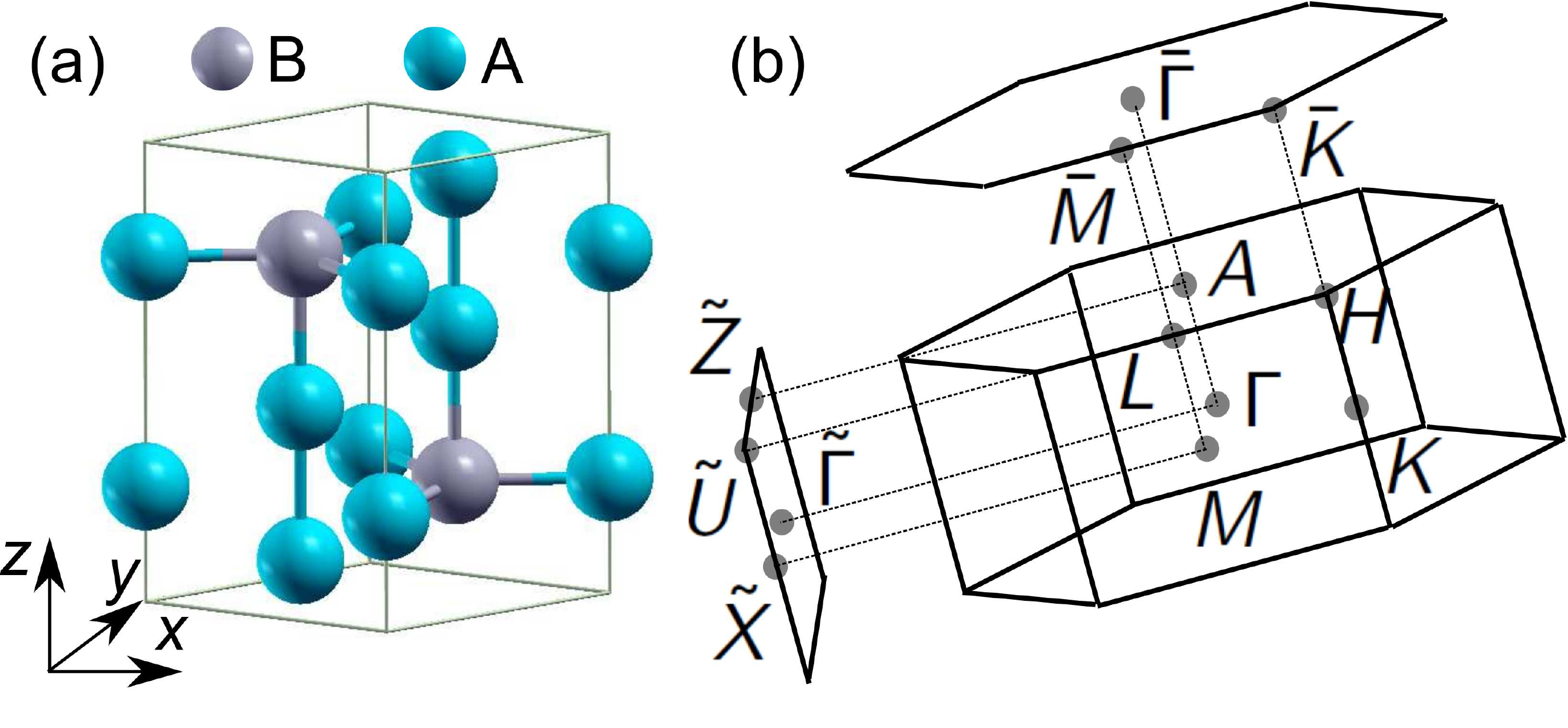}
  \caption{(Color online) (a) Hexagonal unit cell for A$_3$B compounds, with
  A=Na, K, Rb and B=Bi, Sb. (b) Bulk and surface projected Brillouin zone for
  the structure with the high symmetry points marked. The three dimensional
  Dirac crossing occurs along the $\Gamma -A$ direction.} \label{prelim}
\end{center}
\end{figure}

\textit{Computational Methods}. We have carried out first-principles calculations by using the projector augmented 
plane wave method as implemented in Vienna Ab-initio Simulation Package (VASP)~\cite{vasp}, and employed the
Perdew-Burke-Ernzerhof parameterization of the exchange-correlation functional~\cite{pbe}. Spin orbit coupling was 
included for all computations in a self-consistent manner. The electronic structure simulations have been performed 
with a plane wave cutoff of 600~eV on a $8\times 8\times4$ Monkhorst-Pack $k$-point mesh. All the A$_3$B compounds 
(A=Na, K, Rb, B=Bi, Sb) investigated here crystallize in the $D^{4}_{6h}$ structure, as shown in Fig.~\ref{prelim}. During 
the structural optimization the atomic coordinates were allowed to relax until total energy differences were less than 1~meV. 
From the bulk first-principles results, we have projected onto a basis of Na $3s$ and Bi $6p$ (Sb $5p$) orbitals by using a 
procedure based on constructing Wannier functions~\cite{wannier90}. The obtained tight-binding parameters were then 
used to study slab geometries. By combining this scheme with a coherent potential approximation (CPA) including 
self-energy corrections for disorder, we have investigated the \nabisb~alloy~\cite{soven-cpa}. We note that this methodology 
has been recently used to predict the robustness of Dirac fermions in Topological Crystalline Insulator (TCI) alloys, as well 
as in Ferroelectric Rashba Semiconductor (FERSC) alloys~\cite{picozzi-cpa}.
\begin{figure}[t]
\begin{center}
  \includegraphics[scale=0.38]{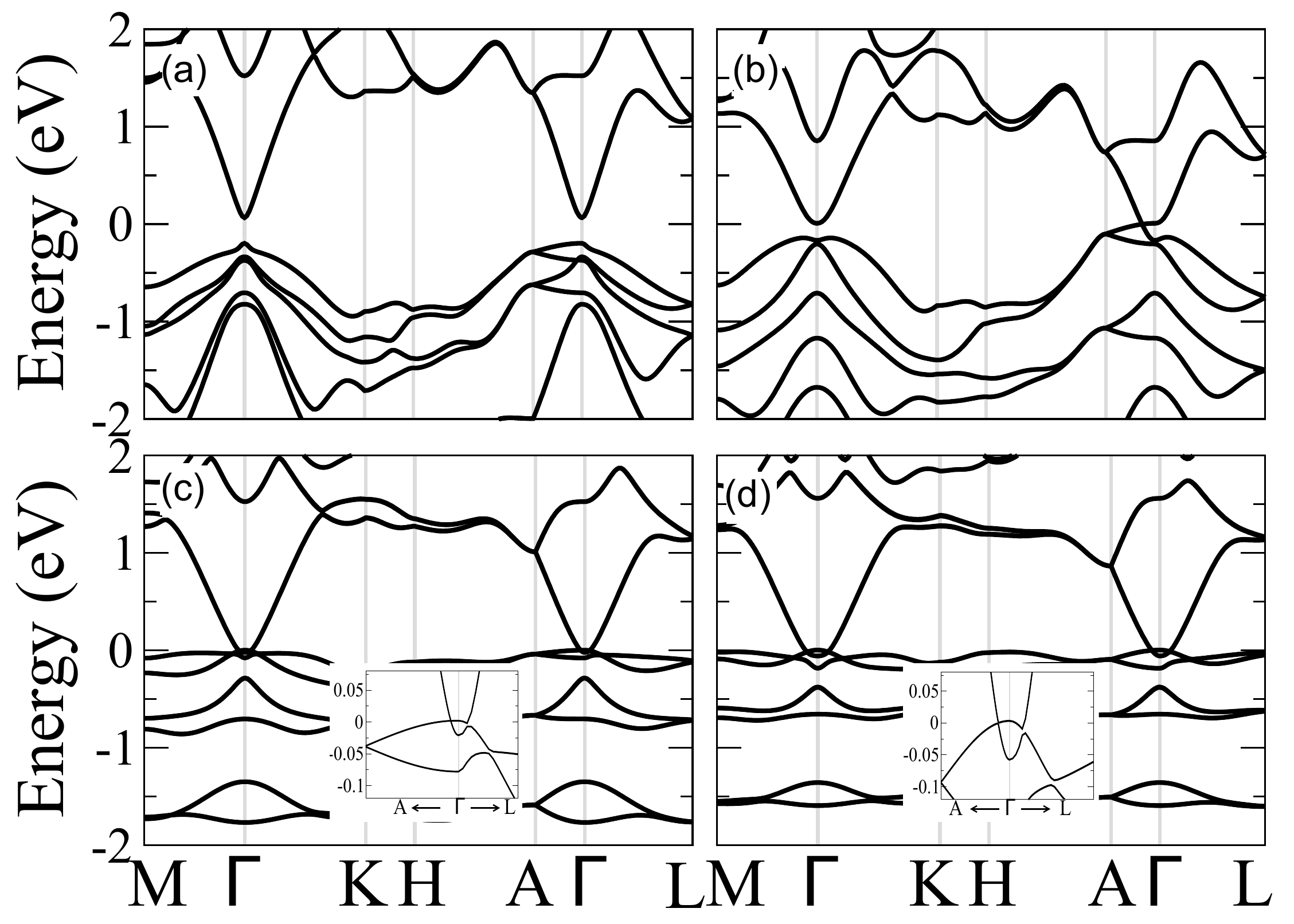}
  \caption{Bulk band structures obtained by including spin orbit interaction for (a) Na$_3$Sb, (b)
  Na$_3$Bi, (c) K$_3$Bi and (d) Rb$_3$Bi. Note the Dirac crossing in (b)-(d).
  The insets in (c) and (d) show a zoom around $\Gamma$ with the crossing along
  $\Gamma -A$.} \label{bulk_bands}
\end{center}
\end{figure}

\textit{Results and discussions}. We begin our analysis by calculating the relativistic bulk band structures for the four 
materials Na$_3$Sb, Na$_3$Bi, K$_3$Bi and Rb$_3$Bi, as shown in Fig.~\ref{bulk_bands}. For Na$_3$Bi we find
the three dimensional Dirac crossing along the $\Gamma-A$ line, and a band inversion at the BZ center, which is 
consistent with the previous study of Wang \textit{et al.}~\cite{fang-na3bi}. Na$_3$Sb, in contrast, is a small gap
insulator with a conventional band ordering. Our calculations reveal that on replacing Na in Na$_3$Bi with heavier 
atoms, the resulting compounds K$_3$Bi and Rb$_3$Bi are metallic with small electron pockets around $\Gamma$. 
However the crossing away from $\Gamma$ is still present. The band structures for the two materials are shown in 
Fig.~\ref{bulk_bands}(c) and (d), along with a zoom around the BZ center in the insets. 
A transition from a hexagonal to a cubic form has been reported for K$_3$Bi and Rb$_3$Bi at high temperatures~\cite{sands,chuntonov}. 
So, our results for the band structures of these two materials are expected to hold at low temperatures.
\begin{figure}[b]
\begin{center}
  \includegraphics[scale=0.45]{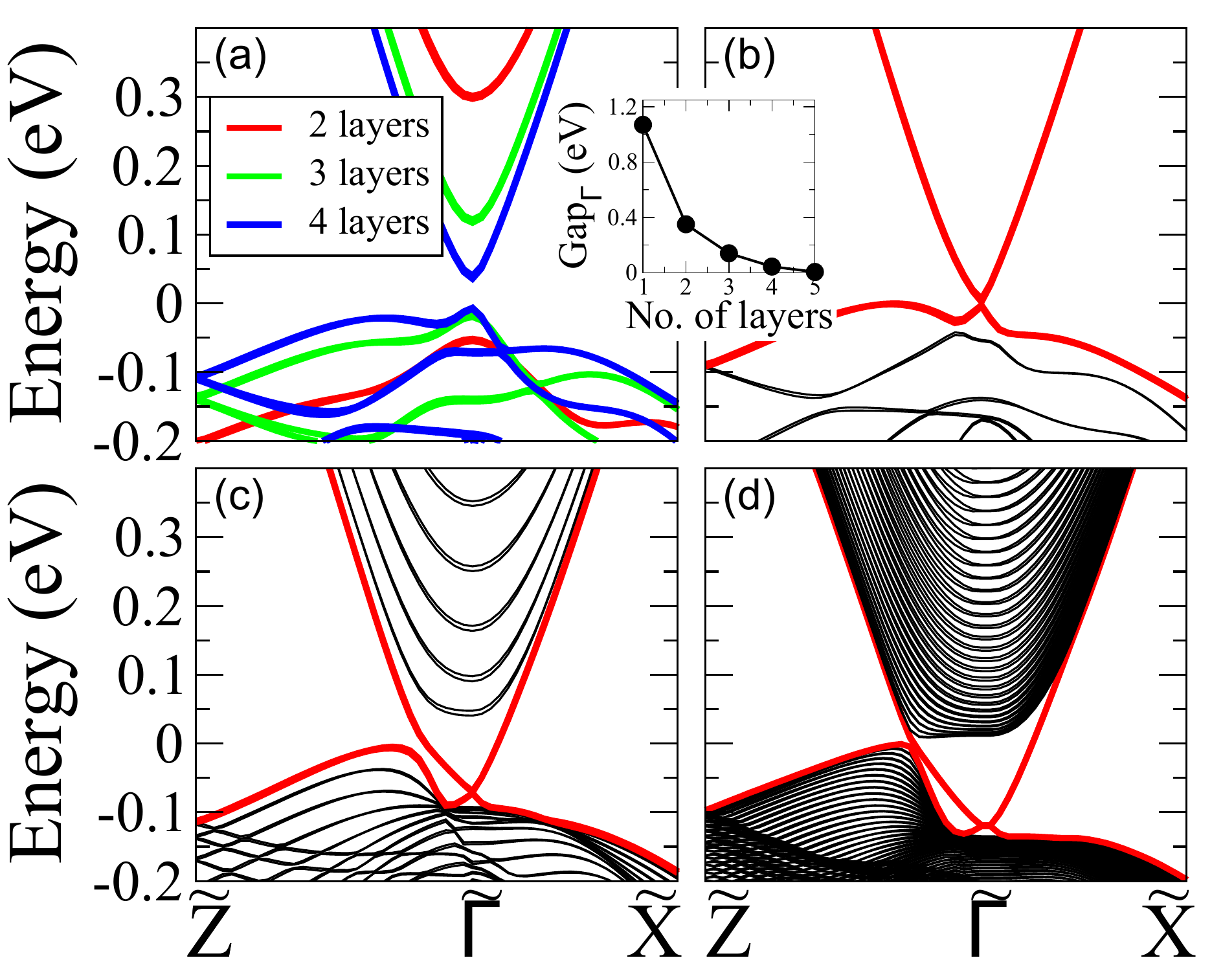}
  \caption{(Color online) Band structures for Na$_3$Bi thin films of thickness
  (a) 2-4 layers, (b) 5 layers, (c) 20 layers and (d) 100 layers. Inset in
  (a)-(b) shows the energy gap at the center of the Brillouin zone for slabs of
  thickness 1 to 5 layers. In (b)-(d) Dirac crossings are highlighted in red.} \label{slab_bands}
\end{center}
\end{figure}

\begin{figure*}[t]
\begin{center}
  \includegraphics[scale=1.5]{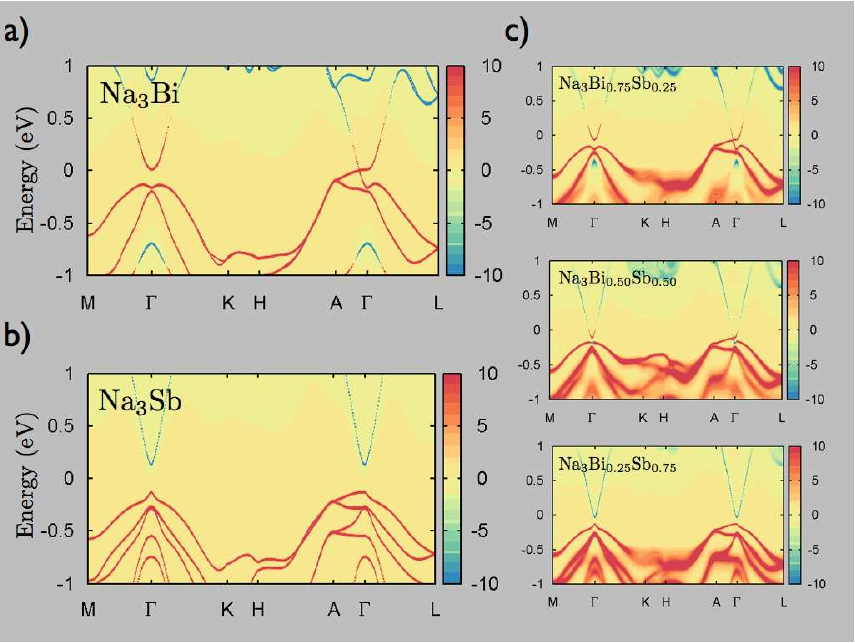}
  \caption{(Color online) Spectral functions for pristine (a) Na$_3$Bi and (b)
  Na$_3$Sb. (c) Spectral functions for the \nabisb~alloy with increasing Sb
  concentration ($x=0.25,0.50,0.75$ from top to bottom). The color scale shows
  the orbital contribution, with red (positive values) denoting Bi/Sb $p$
  orbitals and blue (negative values) representing Na $s$ orbital (in units of
  states/eV).} \label{cpa}
\end{center}
\end{figure*}

Since Na$_3$Bi also shows an inverted band character around the Fermi level, one expects it to form surface states 
when confined into a two-dimensional geometry, similar to a topological insulator. Therefore, we study the evolution 
of the spectrum of Na$_3$Bi films oriented along the $[010]$ surface, as a function of their thickness. For thicknesses 
ranging from 1 to 4 layers, the films are gapped due to an interaction between the two surfaces, as shown in 
Fig.~\ref{slab_bands}(a). This gap decreases monotonically, with the surface cone at $\widetilde{\Gamma}$ becoming 
gapless for a 5-layer-thick film. One can also notice the shoulder along the $\widetilde{\Gamma}-\widetilde{Z}$ direction, 
which rises upwards in energy to form the bulk Dirac crossing for thicker films. This bulk crossing is fully formed only for 
film thicknesses larger than 100 layers ($\approx$ 90~nm). We note that the Fermi arcs can exist with the bulk Dirac nodes, as long as the two-dimensional $\mathbb{Z}_{2}$ invariant on the $k_{z}=0$ plane, $\nu_{\mathrm{2D}}$, is non-trivial~\cite{nagaosa}. Our predictions for the thickness dependence of the surface 
and bulk Dirac cones call for verifications by angle resolved photoemission experiments. Indeed, such measurements 
with varying film thickness have been recently undertaken for topological insulator films~\cite{xue-thinfilmTI,hasan-thinfilmTI}. 
In the case of Na$_3$Bi [010] slabs, one should be able to see two gap-closing transitions at very different film thicknesses: 
one for the surface cone for a few layers slab, with the next gap-closing occurring in the bulk for a hundred layers slab. 
For the [001] surface, the bulk and surface Dirac cones are all projected onto the
Brillouin zone center. In contrast, these Dirac crossings are separated in
reciprocal space for the [010] surface. We decided to focus on the latter, as in this
case it may be easier to distinguish the two cones in a angle-resolved
photoemission measurement. 
Very recently thin films of Na$_3$Bi have been grown by molecular beam epitaxy~\cite{chen-na3bithinfilm}, a development 
which provides a clear route to verify our predictions.

We turn now our attention to the \nabisb~alloy. From the bulk band structures in Fig.~\ref{bulk_bands}, we observe 
that Na$_3$Sb is topologically trivial, having neither the bulk Dirac crossing nor a band inversion at the BZ center, as
opposed to Na$_3$Bi. This opens up the intriguing possibility to obtain a quantum phase transition in \nabisb~solid 
solutions. To this end, we have performed DFT+CPA calculations for the alloy. The spectral functions at different Sb 
concentrations are shown in Fig.~\ref{cpa}. With increasing Sb concentration, the bulk Dirac crossing along $\Gamma-A$ 
moves towards the BZ center. At around a critical concentration of $x_{c}=0.5$ (Na$_3$Bi$_{0.5}$Sb$_{0.5}$), this 
crossing reaches close to $\Gamma$. Upon subsequent increase in Sb concentration, an energy gap appears. We
note that this is the consequence of the annihilation between this Dirac cone and its time reversed partner along the 
$\Gamma$ to $A$ direction. The Sb concentration therefore represents an efficient tool to manipulate the position 
of the bulk Dirac points in $k$-space along the $\Gamma-A$ line. Interestingly, the disappearance of the bulk cone 
is accompanied by a loss of the inverted band character, as it can be evidenced from the reversal in orbital character 
of the valence and conduction bands, before and after passing through the critical Sb concentration. From bulk-boundary 
correspondence, one can then infer that for slabs made of these alloys there would also be a transition in the surface 
spectrum: below $x_{c}$ the surface would display a Dirac crossing, while increasing Sb concentration beyond this value 
would lead to the opening of a trivial gap. Thus, our calculations reveal a topological phase transition in the prototypical 
three-dimensional Dirac semimetal. 

Recently, such tunable phase transitions were experimentally reported for topological insulators and topological crystalline
insulators~\cite{hasan-bitl,ando-bitl,story-tci,hasan-tci}. This makes us confident that our predictions can be verified in the 
near future. It is also worthy considering that our DFT+CPA calculations reveal a protection against substitutional disorder 
of the spectral features of three-dimensional Dirac semimetals around the Fermi level. We note, in fact, the absence of 
broadening of spectral features around the cone, as compared to other energies. Such a robustness, similar to what
happens for topological crystalline insulators and Weyl fermion systems, arises from the three-dimensional nature of the 
Dirac cone~\cite{picozzi-cpa}, and in turn leads to the concrete possibility of experimental verifications by means of
spectroscopic techniques. As shown in Ref.~\cite{picozzi-cpa}, this is a consequence of a vanishing disorder self-energy 
around the crossing point. We also propose that a similar phase transition, and a similar robustness against disorder, 
would occur in the Cd$_3$(As$_{1-x}$P$_{x}$)$_2$ alloy, since the parent compounds Cd$_3$As$_2$ and Cd$_3$P$_2$ 
are Dirac semi-metal and conventional insulator, respectively, with the former having an inverted band order and the latter
having a normal band sequence (see supplemental material~\cite{supplement}).

\textit{Conclusions}. In summary, we have studied the interplay between bulk and surface Dirac fermions in prototypical 
three dimensional Dirac semimetals, by using first-principles-based tight-binding calculations. Furthermore, by means of
density functional theory with coherent potential approximation simulations, we have revealed a topological phase transition 
in \nabisb and Cd$_3$[As$_{1-x}$P$_x$]$_2$. The change of Sb or P concentration provides an efficient way to engineer the reciprocal space position of the three
dimensional Dirac cone, with potential implications for technological devices benefiting from this additional degree of freedom.
Intriguingly, the phase transition from a Dirac semimetal to an insulator is accompanied by a change in the bulk band ordering. 
This can be related, via the bulk-boundary correspondence, to a concomitant transition in the surface state spectrum.

\textit{Acknowledgments}. AN thanks the Irish Research Council (IRC) for financial support. DDS and SP thank the CARIPLO
Foundation through the MAGISTER project Rif. 2013-0726. We acknowledge CINECA, Irish Centre for High-End Computing 
(ICHEC) and Trinity Center for High Performance Computing (TCHPC) for providing computational resources. DDS
and AN thank Prof. C.~Franchini for useful correspondence. DDS kindly thanks Prof. S.~Ciuchi for valuable discussions 
about CPA calculations and disorder, and E. Plekhanov for useful cross-checkings of slab calculations.

\newpage{}

\begin{figure}
 \centering 
 \includegraphics[page={1},width=\textwidth]{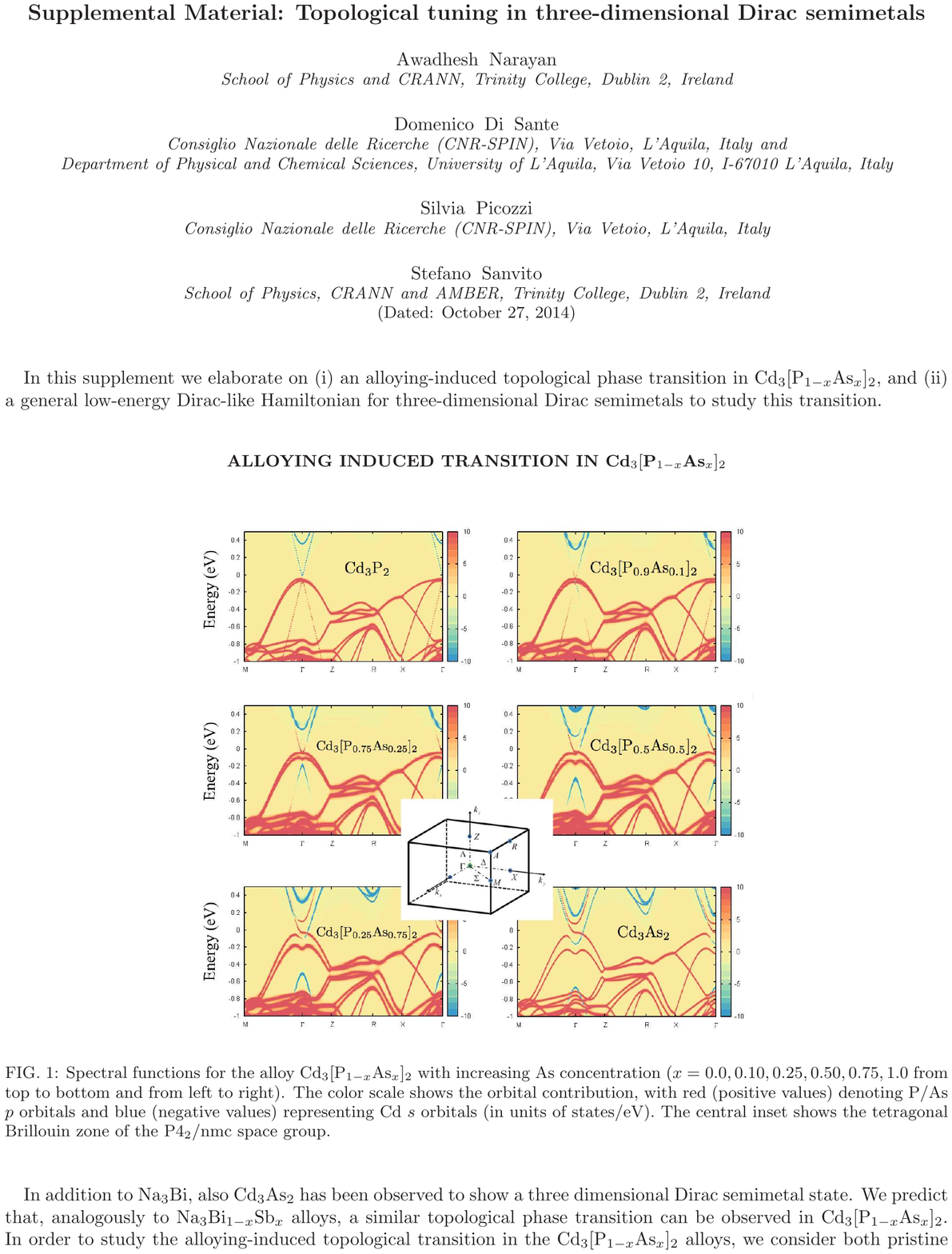}
\end{figure}
 \begin{figure}
  \centering 
  \includegraphics[page={2},width=\textwidth]{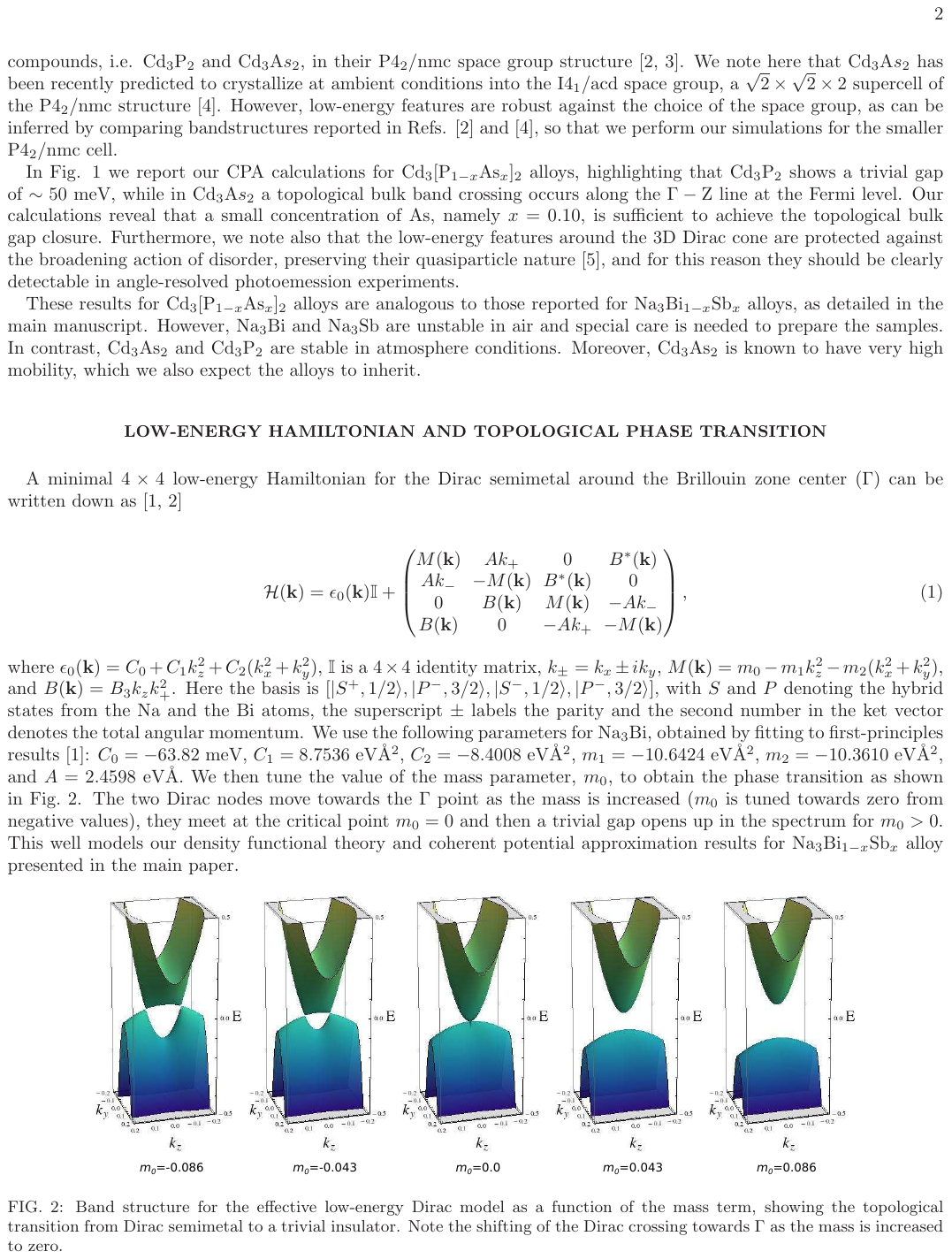}
 \end{figure}
 \begin{figure}
  \centering 
  \includegraphics[page={3},width=\textwidth]{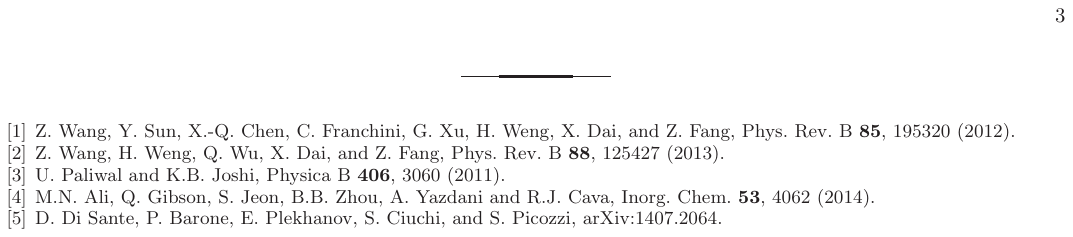}
 \end{figure}


\begin{thebibliography}{99}

\bibitem{review-geim} A.K.~Geim and K.S.~Novoselov, Nature Mater. {\bf 6}, 183 (2007).

\bibitem{review-kane} M.Z.~Hasan and C.L.~Kane, Rev. Mod. Phys. {\bf 82}, 3045 (2010).

\bibitem{review-zhang} X.-L.~Qi and S.-C.~Zhang, Rev. Mod. Phys. {\bf 83}, 1057 (2011).

\bibitem{kane-semimetal} S.M.~Young, S.~Zaheer, J.C.Y.~Teo, C.L.~Kane, E.J.~Mele, and A.M.~Rappe, Phys. Rev. Lett. {\bf 108}, 140405 (2012).

\bibitem{fang-na3bi} Z.~Wang, Y.~Sun, X.-Q.~Chen, C.~Franchini, G.~Xu, H.~Weng, X.~Dai, and Z.~Fang, Phys. Rev. B {\bf 85}, 195320 (2012).

\bibitem{fang-cd3as2} Z.~Wang, H.~Weng, Q.~Wu, X.~Dai, and Z.~Fang, Phys. Rev. B {\bf 88}, 125427 (2013).

\bibitem{chen-na3bi} Z.K.~Liu, B.~Zhou, Y.~Zhang, Z.J.~Wang, H.M.~Weng, D.~Prabhakaran, S.-K.~Mo, Z.X.~Shen, Z.~Fang, X.~Dai, Z.~Hussain, and Y.L.~Chen, Science {\bf 343}, 864 (2014).

\bibitem{hasan-na3bi} S.-Y.~Xu, C.~Liu, S.K.~Kushwaha, T.-R.~Chang, J.W.~Krizan, R.~Sankar, C.M.~Polley, J.~Adell, T.~Balasubramanian, K.~Miyamoto, N.~Alidoust, G.~Bian, M.~Neupane, I.~Belopolski, H.-T.~Jeng, C.-Y.~Huang, W.-F.~Tsai, H.~Lin, F.C.~Chou, T.~Okuda, A.~Bansil, R.J.~Cava, and M.Z.~Hasan, arXiv:1312.7624.

\bibitem{chen-cd3as2} Z.K.~Liu, J.~Jiang, B.~Zhou, Z.J.~Wang, Y.~Zhang, H.M.~Weng, D.~Prabhakaran, S.-K.~Mo, H.~Peng, P.~Dudin, T.~Kim, M.~Hoesch, Z.~Fang, X.~Dai, Z.X.~Shen, D.L.~Feng, Z.~Hussain, and Y.L.~Chen, Nature Mat. {\bf 13}, 677 (2014).

\bibitem{hasan-cd3as2} M.~Neupane, S.-Y.~Xu, R.~Sankar, N.~Alidoust, G.~Bian, C.~Liu, I.~Belopolski, T.-R.~Chang, H.-T.~Jeng, H.~Lin, A.~Bansil, F.C.~Chou, and M.Z.~Hasan, Nature Commun. {\bf 5}, 3786 (2014).

\bibitem{cava-cd3as2} S.~Borisenko, Q.~Gibson, D.~Evtushinsky, V.~Zabolotnyy, B.~Buchner, and R.J.~Cava, Phys. Rev. Lett. {\bf 113}, 027603 (2014).

\bibitem{potemski-znb} M.~Orlita, D.M.~Basko, M.S.~Zholudev, F.~Teppe, W.~Knap, V.I.~Gavrilenko, N.N.~Mikhailov, S.A.~Dvoretskii, P.~Neugebauer, C.~Faugeras, A-L.~Barra, G.~Martinez, and M.~Potemski, Nature Phys. {\bf 10}, 233 (2014).

\bibitem{zhang-bi2se3} H.~Zhang, C.-X.~Liu, X.-L.~Qi, X.~Dai, Z.~Fang, and S.-C.~Zhang, Nature Phys. {\bf 5}, 438 (2009).

\bibitem{damascelli} A.~Damascelli, Physica Scripta. T109, {\bf 61} (2004).

\bibitem{vasp} G.~Kresse and J.~Furthmuller, Comput. Mater. Sci. {\bf 6}, 15 (1996).

\bibitem{pbe} J.P.~Perdew, K.~Burke, and M.~Ernzerhof, Phys. Rev. Lett. {\bf 77}, 3865 (1996).

\bibitem{wannier90} A.~A. Mostofi, J.~R. Yates, Y.-S. Lee, I. Souza, D. Vanderbilt and N. Marzari, Comput. Phys. Commun. {\bf 178}, 685 (2008).

\bibitem{soven-cpa} P.~Soven, Phys. Rev. {\bf 156}, 809 (1967).

\bibitem{picozzi-cpa} D.~Di Sante, P.~Barone, E.~Plekhanov, S.~Ciuchi, and S.~Picozzi, arXiv:1407.2064.

%

\bibitem{sands} Sands \textit{et al.}, J. : Acta. Crystallogr. {\bf 16}, 316 (1963).

\bibitem{chuntonov} Chuntonov \textit{et al.}, Sov. Phys. Crystallogr. {\bf 22}, 367 (1977).

\bibitem{nagaosa} B.-J.~Yang and N.~Nagaosa, Nat. Commun. {\bf 5}, 4898 (2014).



\bibitem{xue-thinfilmTI} Y.~Zhang \textit{et al.}, Nature Phys. {\bf 6}, 584 (2010).

\bibitem{hasan-thinfilmTI} M.~Neupane \textit{et al.}, Nature Commun. {\bf 5}, 3841 (2014).

\bibitem{chen-na3bithinfilm} Y.~Zhang \textit{et al.}, Appl. Phys. Lett. {\bf 105}, 031901 (2014).


\bibitem{hasan-bitl} S.-Y.~Xu, Y.~Xia, L.A.~Wray, S.~Jia, F.~Meier, J.H.~Dil, J.~Osterwalder, B.~Slomski, A.~Bansil, H.~Lin, R.J.~Cava, and M.Z.~Hasan, Science {\bf 332}, 560 (2011).

\bibitem{ando-bitl} T.~Sato, K.~Segawa, K.~Kosaka, S.~Souma, K.~Nakayama, K.~Eto, T.~Minami, Y.~Ando, and T.~Takahashi, Nature Phys. {\bf 7}, 840 (2011).


\bibitem{story-tci} P.~Dziawa \textit{et al.}, Nature Mater. {\bf 11}, 1023 (2012).

\bibitem{hasan-tci} S.-Y.~Xu \textit{et al.}, Nature Commun. {\bf 3}, 1192 (2012).

\bibitem{supplement} See supplemental material for CPA calculations for the Cd$_3$(As$_{1-x}$P$_{x}$)$_2$ alloy, along with a low-energy model to study the topological phase transition.

\end{thebibliography}
\end{document}